\documentstyle[emulateapj,epsf]{article}

\def\l{\ell}
\def\leff{\ell_{eff}}
\def\Dl{\Delta\ell}
\def\pp{\parshape 2 0truecm 8.8truecm 0.5truecm 8.5truecm}
\def\fnote#1#2{$^#1$\pp{#2}}

\slugcomment{Submitted to Astrophysical Journal
Letters August 5, 1998}

\lefthead{Herbig et al.}
\righthead{QMAP 2}

\begin{document}

\title{Mapping the CMB II:
The Second Flight of the QMAP Experiment}

\author{T. Herbig\altaffilmark{1,4}, A. de
Oliveira-Costa\altaffilmark{1,2}, M. J. Devlin\altaffilmark{3},
A. D. Miller\altaffilmark{1}, L. Page\altaffilmark{1}, and
M.~Tegmark\altaffilmark{2,4}}

% \altaffiltext{1}{Princeton University, Physics Department, Jadwin Hall,
% Princeton, NJ 08544; herbig@pupgg.princeton.edu} 
% \altaffiltext{2}{Institute for Advanced Study, Olden Lane, Princeton, NJ
% 08540}
% \altaffiltext{3}{University of Pennsylvania, Department of Physics and
% Astronomy, David Rittenhouse Laboratory, Philadelphia, PA 19104} 
% \altaffiltext{4}{Hubble Fellow}

\begin{abstract}

We report the results from the second flight of {\sl QMAP}\null,
an experiment to map the cosmic microwave background near the
North Celestial Pole.  We present maps of the sky at 31~and
42\,GHz as well as a measurement of the angular power
spectrum covering $40 \la\ell\la 200$.  Anisotropy is detected at
$\ga 20\sigma$ and is in agreement with previous results at these
angular scales.  We also report details of the data reduction
and analysis techniques which were used for both flights of {\sl QMAP}\null.

\end{abstract}


\section{Introduction}

The {\sl QMAP} experiment is a balloon-borne telescope designed to map
the cosmic microwave background (CMB) from angular scales of 0\fdg7 to
several degrees.  Its design is similar to that of the Saskatoon experiment
(\cite{Wollack97}), but the absence of significant atmospheric emission
at balloon altitudes and the interlocking scan pattern allow the
production of a sky map in an area close to the North Celestial Pole
\hbox{(NCP)}.  {\sl QMAP} flew twice in 1996.

The telescope has 6~radiometry channels in 3~dual-polarization beams.
Two of these channels are at Ka-band (31\,GHz) and four are at Q-band
(42\,GHz).  The telescope can be pointed in azimuth, but the beams have
a fixed elevation of about 41\arcdeg.  The three beams are chopped
sinusoidally in azimuth at 4.6\,Hz over an angle of up to $\theta_c =
20\arcdeg$ on the sky.  In addition, the azimuth of the telescope is
``wobbled'' sinusoidally with periods of 50--100\,s and amplitudes of
approximately half the chop amplitude.

This {\it Letter} reports the results from the second flight of {\sl
\hbox{QMAP}}.  It also details the calibration and pointing analysis
common to both flights of the experiment.  A companion paper
(\cite{d98}, hereafter D98) presents the results of flight~1 and
discusses the design and performance of the instrument.  Another
companion paper (\cite{o98}; dO98) presents the combined results from
both flights and explains the mapping technique and the method used to
calculate the angular power spectrum of the CMB.

\section{Observations and Data}
\label{data}

The second flight of the {\sl QMAP} experiment was launched 1996
November 10 at 23:05\,UT from Ft.~Sumner, \hbox{NM}, by the National
Scientific Balloon Facility (NSBF).  The gondola reached its float
altitude of 30\,km two hours 
later and stayed at float until
12:55\,\hbox{UT}.  Science and calibration data were taken betweem 01:00
and 12:10\,\hbox{UT}, all during the night.  The flight was terminated
350\,km east,
near Shamrock, TX\null.

\bigskip

{\footnotesize

\fnote{1}{Princeton University, Physics Department, Jadwin Hall,\\
Princeton, NJ 08544; herbig@pupgg.princeton.edu}

\vskip0.1cm
 
\fnote{2}{Institute for Advanced Study, Olden Lane, Princeton, NJ 08540}

\vskip0.1cm

\fnote{3}{University of Pennsylvania, Department of Physics and\\
Astronomy, David Rittenhouse Lab., Philadelphia, PA 19104}

\vskip0.1cm
 
\fnote{4}{Hubble Fellow}
}

\goodbreak

\noindent

In contrast to flight~1, this flight was optimized for a higher
signal/noise ratio for each map pixel; this was achieved with a smaller
chop of $\theta_c = 5\arcdeg$ and a wobble of $\Delta \hbox{\it az} \approx
3\arcdeg$ at a period of 50\,s.  Because the latitude of the flight was
$\approx 3\arcdeg$ higher than flight~1, the resulting maps are
closer to the NCP\null.  The sky coverage for both flights is shown in
D98.
%%% LET THIS BEGIN NEW PARAGRAPH IF SPACE PERMITS:
The target region around the NCP was observed in 3~segments of 17305\,s,
9200\,s, and 3974\,s duration, interrupted by celestial calibrations.
At the chop frequency of 4.6\,Hz and the data rate of 160~samples/chop,
% $2.24\times10^7$ 
22,433,920
radiometry samples on the NCP were gathered in each
channel.  All 6~channels functioned cleanly---no data had to be
excised because of excess noise.

During ascent, the attitude computer froze repeatedly, which delayed
starting the radiometer cooling system and prevented the radiometer from
reaching thermal equilibrium.  The resulting amplifier gain variations
were tracked and corrected using periodic calibration pulses.  Also
during ascent, the data acquisition computer froze for $\approx 50$\,s
but resumed functioning without rebooting.  Consequently, the data
buffers overflowed, resulting in large timing offsets between
radiometry, chopper, and pointing samples.  This complicated the phase
solutions described in section~\ref{phases}.  No other timing delays
were encountered.

The primary CCD pointing camera had an 8\arcdeg\ field of view, while
the secondary camera covered 12\arcdeg; both guided with $<10\arcsec$
intrinsic pointing noise at 10\,Hz.  Almost all observations were made
using the narrow-field camera.  As discussed in section~\ref{pointing},
no stars were observed in the CCD during the radiometer calibration on
Cas\,A, resulting in a large azimuth pointing uncertainty.  Two ballast
drops during the second NCP segment, tracked well in the pointing, had
no discernible effect on the data.

% The offset transfer between cameras is
% better than 0\farcm3 and 1\farcm2 in elevation and azimuth,
% respectively.

Apart from the radiometer, all system components reached thermal
regulation quickly.  Exposed passive elements, such as the chopper and
the mirror, cooled to temperatures of $\approx -45$\,C upon reaching
float altitude, and stayed within 10\,K for the remainder of the flight.
The gondola position was recorded every minute by the GPS receivers in
the NSBF flight package; time was recorded to 2\,s absolute accuracy
with the on-board clock.

% {\sl
% The inclinometers indicate that the long-term tilt of the telescope
% (balanced before launch) remained constant to within 0\fdg1g...
% }

\section{Analysis}

The analysis procedure is similar for both flights of {\sl QMAP}; the
key differences are a degraded CCD focus in flight~1 and large phase
delays in flight~2.

% The analysis uses the on-board data at full time resolution; the
% compressed downlink data are discarded.  Preliminary steps include
% matching the flight log to the on-board history, pointing status, and
% the GPS~log, unwrapping the system synchronization clock, perusing each
% housekeeping channel, and identifying sections of acceptable pointing
% lock.

\subsection{Pointing}
\label{pointing}

For both flights, the CCD~camera was the primary pointing device.  For
each second of data, the {\it az}/{\it el} positions of the guide stars and
planets were calculated using the on-bard clock and the GPS position of
the telescope.  The camera scale was calibrated by measuring the
positions of several stars visible in the camera field while the
telescope was held at constant azimuth using the magnetometer.  From
laboratory measurements of the chopper and camera orientations and our
limits on the camera horizon during flight, we find that the maximum
deviation of the beam at the endpoints of the maximum chop
($\theta_c=20\arcdeg$) is $<0\fdg1$ in elevation.  

% This is not included in the pointing solution.

% The offset of the chopper zero-point is contained in the
% optical-radio offset and is therefore determined automatically.

The offset of the CCD center from the radio beams was determined
directly in flight~1, where a guide star was available while the radio
source Cas\,A was observed.  During flight~2, no guide star was found
(Cas\,A was setting instead of rising), and the radio center (at
$\hbox{\it az}=-39\arcdeg$) had to be transferred to the CCD center (at the NCP)
using the magnetometer.  The magnetometer was calibrated using CCD star
positions at various azimuths ($-0\fdg8 < \hbox{\it az} < +138\fdg1$, which
excludes the azimuth of Cas\,A) and a global fit during a full spin of
the telescope.  From this, we estimate the remaining uncertainty in the
radio positions at $0\fdg5$ in azimuth, or $0\fdg4$ on the sky.  The
resulting smearing of map pixels is minor: depending on the channel, the
average smearing is 0\fdg06, with only 10\%\ of Ka1/2 and Q1/2 samples
smeared by more than 0\fdg12 (0\fdg19 for Q3/4).  No section of the map
is smeared more than 0\fdg2 (0\fdg3 for Q3/4), small compared to the
beam sizes.

% ^^^ these are the final numbers now ^^^

The pointing analysis of flight~1 was complicated by a defocussed
CCD~camera, which resulted in 5\arcmin\ pointing noise at 10\,Hz as well
as distortions of the CCD camera field of $\approx 0\fdg5$ at the edges.
The pointing noise could be smoothed easily, while the field distortions
caused the azimuth wobble to appear as elevation nodding at the wobble
frequency and its first harmonic.  From observations of Cas\,A we found
this nodding to be spurious and suppressed it for subsequent analysis.

Additional pointing effects include elevation drifts at $\tau\sim300$\,s
of $\Delta\hbox{\it el}\sim0\fdg1$ and 0\fdg4 for flights~1 and~2,
respectively, which are likely caused by altitude changes of the
balloon.  In addition, pendulation modes (primarily at 15\,s) were
excited on occasion to amplitudes of 0\fdg1--0\fdg2, with damping times
of 10\,min to half amplitude.  All of these modes were tracked with the
camera in azimuth and elevation and are included in the pointing
solution.  Their effect on telescope roll is of second order and could
not be tracked.

\subsection{Phases}
\label{phases}

The {\sl QMAP} data set consists of three parallel data streams:
radiometry data, chopper position, and CCD star position.  The optimum
phases to associate a radiometry sample with a chopper and a pointing
position are determined by maximizing the signal from Cas\,A with a
two-dimensional search in relative phase of the radiometry to both the
chopper and the pointing. The two levels of pointing modulation (chop
and wobble) are essential to breaking the 180\arcdeg\ degeneracy
encountered with one modulation only.

The phase fits for flight~2 were more complicated than those for
flight~1 because of the buffer overflows described above.  While the
flight~1 chopper-to-pointing lag was 66\,ms and the data-to-chopper lag
was 46--48\,ms, the corresponding lags for flight~2 were 12.17\,s and
19.34--19.40\,s, respectively.  Despite being large, these phases are
accurate to one chopper cycle for pointing and to 0.08~samples (0.1\,ms)
for the data, indicating that there is no significant phase smearing in
either flight.  The calibration pulses gave an independent check of the
phase solutions in both flights.

\subsection{Gain Drift Correction}
\label{pulses}

As described in D98, we inject calibration pulses of constant strength
and roughly 40~samples duration into each radiometry channel before the
first amplifier, allowing us to monitor the total gain of each channel
throughout the flight.
%  Because of their high signal/noise ratio
% (between 20 and 100), the pulses are easily identified and measured.
% Over an entire flight, their start phases drift $< 0.15$~samples.
% start phase: constant to within 0.3 syncs (Ka) and 0.5 syncs (Q) in
%              flight 2.  This makes it 0.125 samples peak-peak.
The pulses reveal that in both flights the instrument gain drifted
smoothly by about 10--15\%\ in each channel.  These drifts were modeled
and applied as corrections in the subsequent analysis.  The corrected
gains vary $< 1$\%\ throughout the flight.

% these numbers come from:
% - sigma of peak fit residuals in flight 2 (for 0.4-2.5%)
% - the fact that there are at least 10 spikes making up each ``segment''
%   of the gain polynomial.  This is likely better, since there are
%   about 100 spikes/polynomial coefficient, so the error would be
%   reduced by a factor of ~10.

\subsection{Radiometry Offsets}
\label{radoff}

The chop-synchronous radiometry offset was determined during mapmaking
as described in dO98.  This offset is produced by the variation in
effective emissivity of the chopper plate (\cite{Wollack97}) and by the
modulation of emissive cavities in the optical layout.  In the mapping
analysis, the radiometry offset was calculated over one chopper cycle
without regard to the actual chopper position.  When the results are
folded by chopper deflection, the offsets at overlapping positions are
in agreement, as illustrated in Figure~\ref{fig:offset} for the Ka1
channel.  This validates both the phasing and the radiometry offset
analysis.

\placefigure{fig:offset}

\medskip
\vbox{
% \centerline{\epsfxsize=8.5cm\epsfbox{Ka1_QMAP12.ps}}
\centerline{\epsfxsize=8.5cm\epsfbox{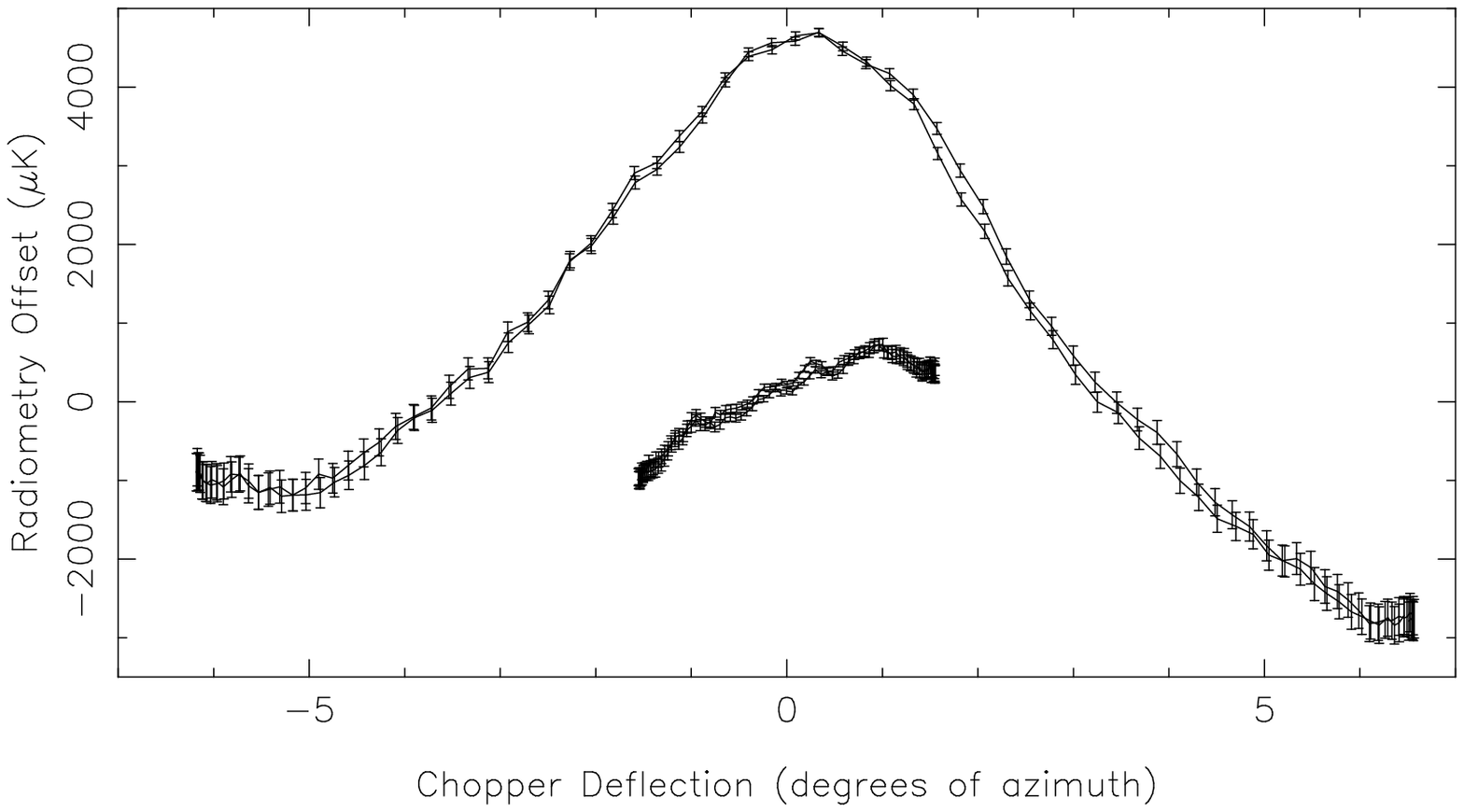}}
\figcaption[herbig_fig1.ps]{Chop-synchronous radiometry offset for
channel Ka1 from flight~1 (wide curve) and flight~2 (narrow curve) as a
function of chopper deflection.  Note the agreement of the folded
curves, which is similar in the other channels.
\label{fig:offset}}}
\smallskip

\subsection{Beams}
\label{beams}

The beam patterns were measured as part of our system calibration, with
$\theta_c = 3\arcdeg$ and $\Delta \hbox{\it az}=5\arcdeg$ in flight~1 and 1\fdg4 in
flight~2.  Because the elevation of the beams is fixed, the vertical
component of the beam map is 
obtained by a drift scan.

The beam maps were calculated by gridding the raw data into 0\fdg1
pixels.  For flight~1, the chop-synchronous radiometry offset discussed
in section~\ref{radoff} was subtracted directly from each channel,
flattening the map background by more than a factor of 10. In flight~2,
the chop amplitude for CMB observations was smaller than that used for
Cas\,A, so that the radiometry offset could not be subtracted from the
beam maps.  Figure~\ref{fig:beam} shows two beam images from flight~1.
The maps for flight~2 are similar, except for slightly larger background
variations.

% All beam maps were further corrected for the background depression
% produced by the AC-coupled detector system, which appears as a
% horizontal band at the elevation of Cas\,A.  Calibration pulses during
% the calibration period were excised.

% Notes to the editors:

\placefigure{fig:beam}
\notetoeditor{Please put the two PS files containing the two beams
side-by-side (K12grey.ps left and Q12grey.ps right).  This will make the
images small, which is intended to avoid drawing undue attention to
them.  I was unable to combine the two images into one PS file.
Thanks.}
 
\medskip
%\vbox{
%\hbox to\hsize{\hbox{\epsfxsize=4.38cm\epsfbox{K12grey.ps}}%
%\hfil\hbox{\epsfxsize=4.38cm\epsfbox{Q12grey.ps}}}
\vbox{
\hbox to\hsize{\hbox{\epsfxsize=4.38cm\epsfbox{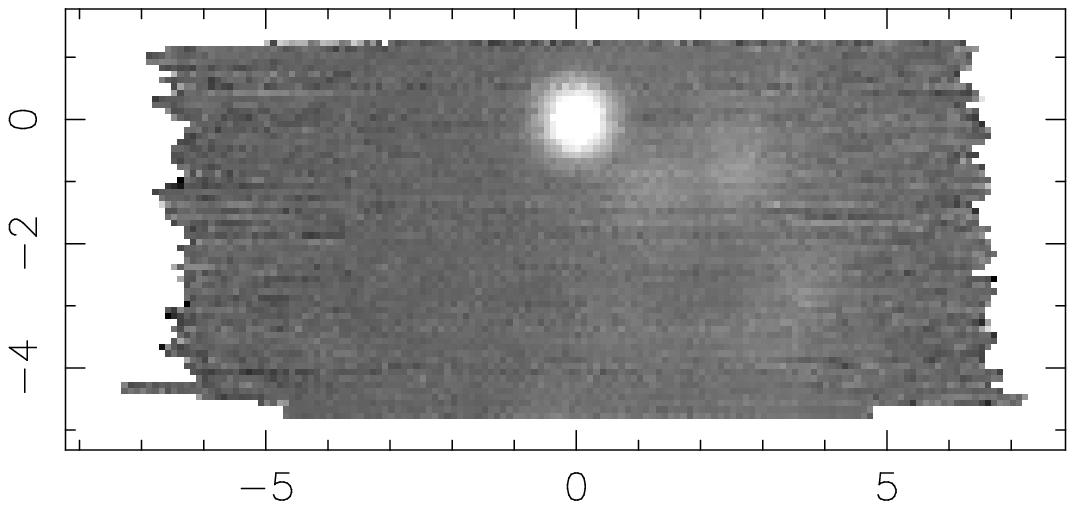}}%
\hfil\hbox{\epsfxsize=4.38cm\epsfbox{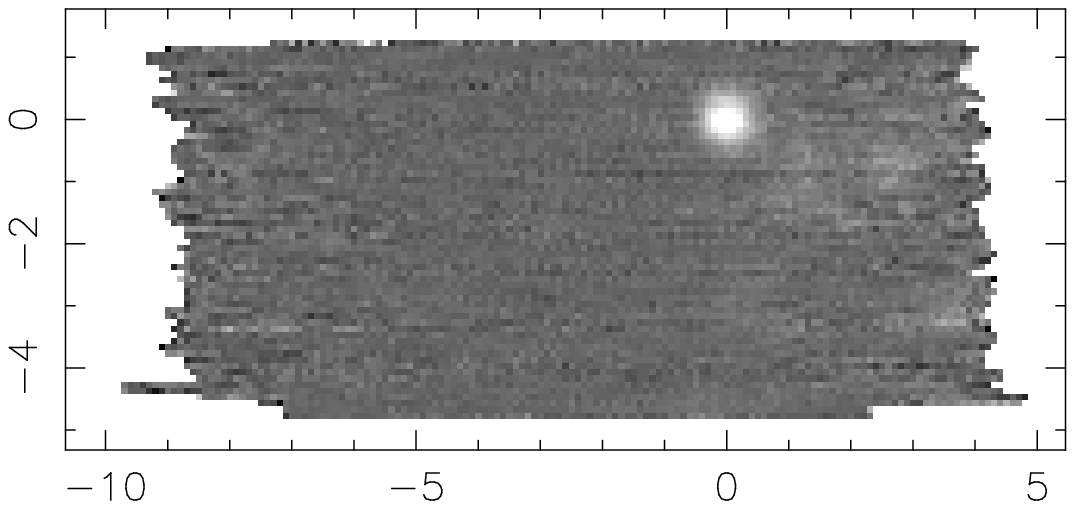}}}
\vskip-0.3cm
\figcaption{
%Beam images of Cas\,A from flight~1, with
%the combined Ka1/2 channels on the left and Q1/2 on the right. 
Flight 1 beam images of Cas\,A from the combined Ka1/2 channels 
(left) Q1/2 channels (right). 
The
chop-synchronous radiometry offset was subtracted.  The grey-scale
values stretch linearly from $-10$ to $+15$\,mK, and the axes are
relative elevation and azimuth in degrees on the sky.  The structure to
the bottom right of Cas\,A corresponds to radio sources in the
Galaxy.\label{fig:beam}} }
\smallskip

For the beam shape determination, corresponding pairs of beams (both
polarizations for each feed, {\it e.g.}, Ka1/2) were combined and fitted
as elliptical Gaussians.  The five shape parameters ({\it az}/{\it el}
position, major and minor FWHM, and angle) for the combined beams were
then used as constraints for the peak fits in each individual channel.
The beam sizes are 
$0\fdg89\pm0\fdg03$,
$0\fdg66\pm0\fdg02$ and
$0\fdg70\pm0\fdg02$
for Ka1/2, Q1/2 and Q3/4, respectively.
The robustness of these fits was explored by varying the fit window size
and the background subtraction model.  The resulting systematic error
estimates in the beam integrals ($\approx 4$--8\%) exceed the formal
errors on the fit parameters and have been included in the overall
calibration errors.

Our susceptibility to errors in the beam sizes is small: for the
calibration, only the error in the beam integral is relevant.  With our
mapping approach, errors in the beam size only affect the highest
multipoles of the angular power spectrum, as discussed in dO98.
The beam parameters were measured independently for each
flight and are in good agreement.

%   --------------------------
%   channel    FWHM difference
%   --------------------------
%   Ka1/2      0.893 +/- 0.029
%   Q1/2       0.660     0.016
%   Q3/4       0.698     0.024
%   --------------------------

\subsection{Calibration}

Our primary calibration source is the supernova remnant Cas\,A,
unresolved at our beam sizes, for which we used the flux densities
compiled by Baars et al.\ (1977) augmented by data points between~31 and
250\,GHz (\cite{leitch,chini,mezger}).  The resulting spectrum fit is
$\log(S_\nu/\hbox{Jy})=(5.713\pm0.023) - (0.759\pm0.006)\log(\nu/\hbox{MHz})$
at epoch 1980.0.  We use the secular decrease model from Baars et al.\
(1977) to predict the flux density at our observing epochs.  The
uncertainty in the flux density of Cas\,A is approximately 8.7\%
(including the uncertainty in our center frequencies of $\approx
0.4$\,GHz) and is the largest single contributor to the overall
calibration uncertainty.

Using the beam fits described in section~\ref{beams}, we predict an
antenna temperature for Cas\,A between 15 and 25\,mK, depending on band
and channel.  Each flight is calibrated independently.  

Note that the beams used for the calibration were observed with the same
strategy as the CMB~data; this means that most stable systematic errors
that could lead to defocussing ({\it e.g.}, pointing noise and drifts,
pendulation, phases, and CCD camera and chopper calibrations) are
automatically corrected to the appropriate calibration.

The total calibration uncertainties are between 10\%\ and 13\%,
depending on channel and flight.  Roughly equal amounts are uncorrelated
(beam fit and center frequency) and correlated (Cas\,A) between
channels.

% {\sl Table for calibration errors?  Integrate with beam parameters table?}

% \subsection{Time Power Spectra}

\subsection{Data Rejection}

A small portion of the data (7.8\%\ and 4.0\%\ for flights 1 and 2,
respectively) were rejected from the analysis.  The bulk (4.8\%\ and
4.0\%) were cut around the calibration pulses (every 100\,s) to allow
recovery of the AC-coupled radiometer baseline.  Smaller cut windows
resulted in a time-dependent radiometry offset, while larger windows
affected neither the radiometry offset nor the final results.  The
remaining rejections were due to pointing glitches (0.7\%\ and 0.02\%)
and to lost guide stars (2.3\%\ in flight~1).  Apart from removing Q1 of
flight~1 from analysis (see D98), there were no cuts based on the
contents or quality of the radiometry data themselves.

% {\sl More cuts: at start of ncp1 in flight~1: lots of star tracking noise,
% fixed by XXX; XXX frames deleted (star close to bottom of frame).}

\section{Results and Discussion}

The techniques for producing sky maps and determining the angular
power spectrum are described in detail in dO98.  The mapmaking
step filters the data at the fundamental and the first harmonic of the
chop frequency to remove the chop-synchronous offset and pre-whitens the
time power spectrum to minimize the noise correlation between samples.
The map is then derived from the linear inversion of a highly
over-determined system of equations involving the filtered data points
and their corresponding sky pixels
and residual scan-synchronous offsets.  
The final map product consists of a
vector ${\bf x}$ of $N$ pixels (each with some position on the sky)
and its $N\times N$ noise covariance matrix ${\bf\Sigma}$.

\placefigure{fig:maps}

\medskip
\vbox{
\centerline{\epsfxsize=8.5cm\epsfbox{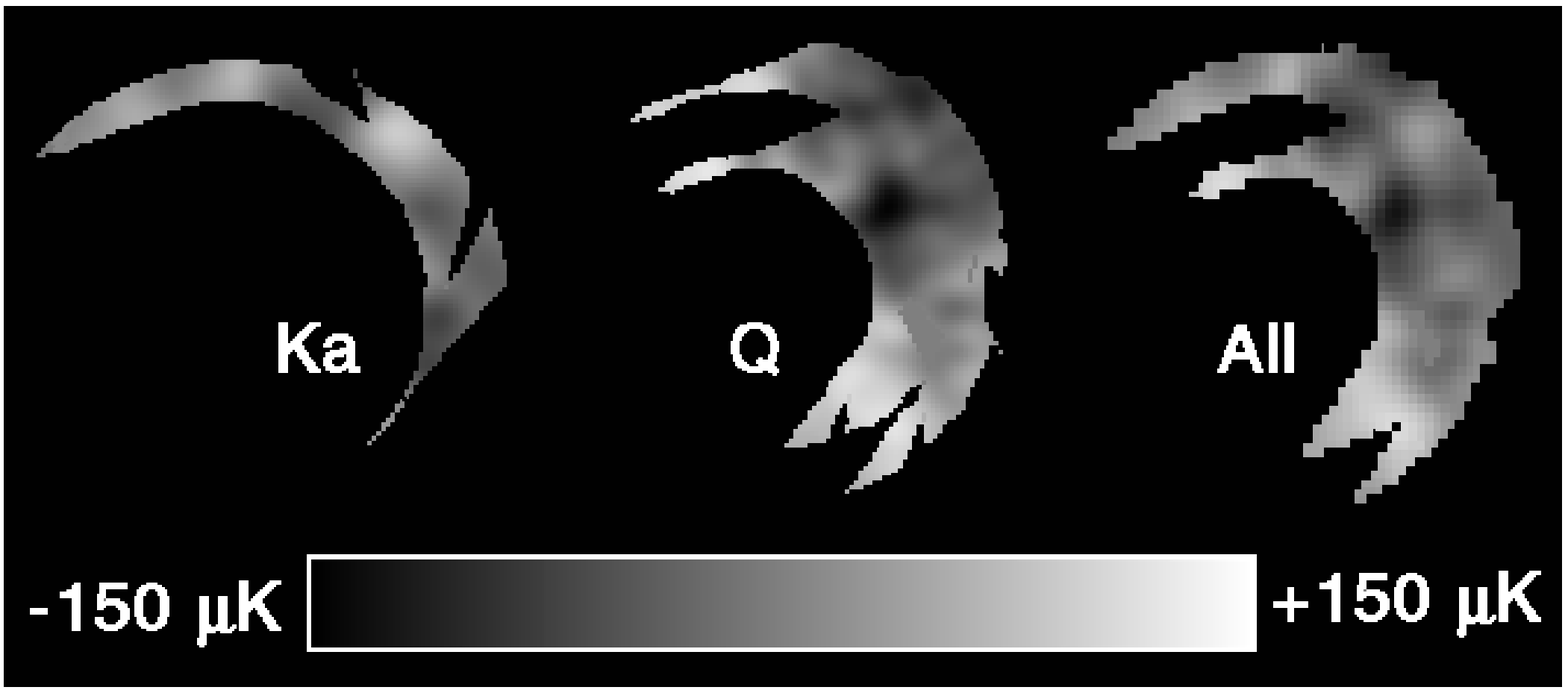}}
\figcaption[maps.ps]{Wiener-filtered maps from flight~2.
The combined Ka-band map (left), the combined Q-band map (middle)
and the combined map from all channels (right)
are shown with the NCP is at the center of the arcs
and ra=0 at the top, increasing clockwise.  The rightmost map
subtends $15\arcdeg\times 16\arcdeg$. These data are
independent of those in D98.
% On the left is the combined Ka-band map; the combined Q-band map is
% shown in the center.  The right-hand map is the combined map from all
% channels and presents all data.  The NCP is at the center of the circle
% and $\alpha=0$ is at the top, increasing clockwise.  The map on the
% right subtends $16\arcdeg\times 15\arcdeg$. These data are
% independent of those in D98.
\label{fig:maps}}}
\smallskip

For visualization, Figure~\ref{fig:maps} shows the Wiener-filtered 
versions of the maps 
(${\bf x}_w\equiv{\bf S}[{\bf S}+{\bf\Sigma}]^{-1}{\bf x}$), 
where ${\bf S}$ is the covariance
matrix due to the observed level of sky fluctuations.  Such filtering
accentuates the statistically significant features of a map. 
% For visualization, the maps are Wiener filtered (${\bf x}_w\equiv{\bf
% S}[{\bf S}+{\bf\Sigma}]^{-1}{\bf x}$, where ${\bf S}$ is the covariance
% matrix due to the average level of sky fluctuations).  Such filtering
% accentuates the statistically significant features of a map.  Combined
% Wiener-filtered sky maps for each band are shown in
% Figure~\ref{fig:maps}, along with a total combined map of all the data
% from flight~2.

The angular power spectrum of the sky in Figure~\ref{fig:pspec} is produced
directly from the combined unfiltered map in each band by expanding in
signal-to-noise eigenmodes and sorting by decreasing signal-to-noise
ratio.  Ranges of eigenmodes are averaged to produce the points shown in
Figure~\ref{fig:pspec} and listed in Table~1.  
They do not contain the
calibration error of 12\%\ for Ka and 11\%\ for Q
\footnote{At the highest $\ell$, the fraction of the calibration
error due to uncertainty in the beam is partially offset by a
change in the window function. 
To be conservative, we do not include this effect}; this error is
systematic and is correlated between the bands of {\sl QMAP} as well as
the Saskatoon results (\cite{cbn97}).

%\medskip
\vbox{
\vskip-1.4cm
\centerline{\epsfxsize=8.5cm\epsfbox{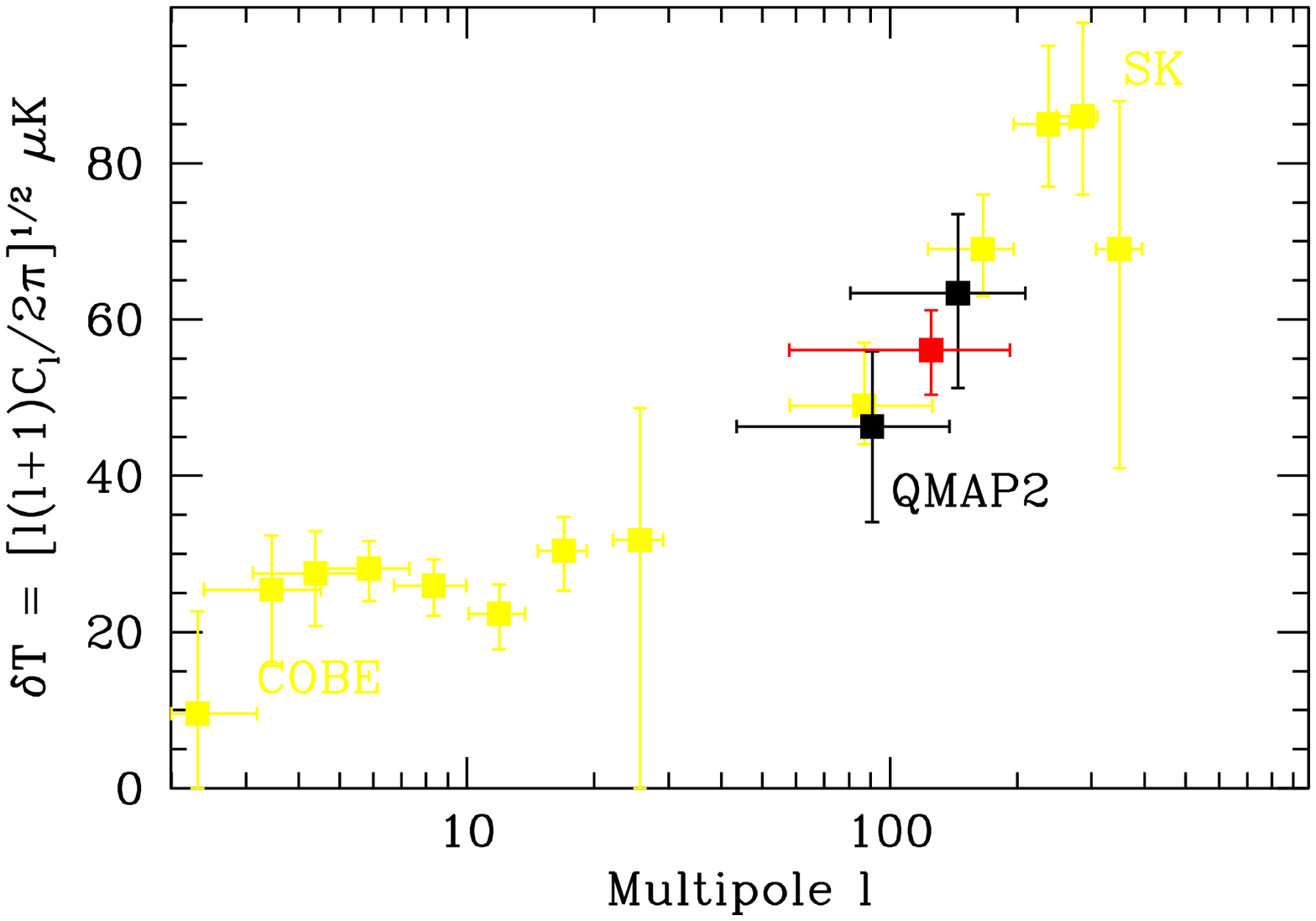}}
\vskip-1.5cm
\figcaption[pspec.ps]{Angular power spectrum from
flight~2.  The central {\sl QMAP} point represents Q-band. The
other two represent Ka-band and have uncorrelated errors.
The error bars include sample variance, but not
the calibration error.  For comparison, we show the results from
{\sl COBE} and Saskatoon (see Tegmark 1996).
\label{fig:pspec}}}
\smallskip

Checks for systematic errors can be made by differencing the maps of
each polarization pair, which cover the same sky pixels because they use
the same feed.  We use a generalized $\chi^2$ statistic (dO98) to
measure the signal-to-noise ratio $\nu$ in each difference map ${\bf
x}_d={\bf x}_a-r{\bf x}_b$, where $a$ and $b$ denote the two maps ({\it
e.g.}, Ka1/2).  When $r=1$, the sky signal is subtracted completely and
$\nu$ should be a minimum.  On the other hand, with $r\ll 1$ 
($r\gg1$), map $a$ (map $b$) 
dominates and $\nu$ gives the significance of the
anisotropy detection in this map.  Figure~\ref{fig:null} shows this
comparison for all three map pairs and illustrates that despite the
substantial signal (which is detected at 
the significance level of 
$3\sigma$, $2\sigma$, $1\sigma$, $6\sigma$, $13\sigma$ and $20\sigma$
in the Ka1, Ka2, Q1, Q2, Q3 and Q4 channels, respectively),
none of the difference maps show any
significant remaining structure. 
Q3 and Q4 are particularly significant since 
they are the only maps to cover the big cold spot seen
in Figure~\ref{fig:maps}
and in the Saskatoon data (Netterfield et al.~1997).
The spectral analysis
of the data will be presented in a future paper.

%The spatial overlap of the Q
%and Ka~maps is too small to constrain the spectral index of the signal
%significantly.

% LP: how about beta? is it really not doable?  If it can be done, is
%     it powerful enough to start messing with the layout?

\placefigure{fig:null}

\section{Conclusions}

The second flight of {\sl QMAP} produced a map of the sky covering
83~square degrees at resolutions of 0\fdg7--0\fdg9 and frequencies of
42--31\,GHz with $\ga 20\sigma$ detection of anisotropy.  
Systematic checks
strongly suggest that this anisotropy is on the sky rather than produced
by the instrument.  We also present an angular power spectrum
covering $40<\ell<200$ produced from the map.  In this region of the
sky, the CMB is the dominant source of anisotropy 
at 30-40 GHz (\cite{dOC97}); thus
we interpret the observed signal as predominantly CMB anisotropy.
The raw data will be made public
upon publication of this {\it Letter}.

% LP: remark about window functions and calibration errors here?
% LP: should we have a remark about how much sample variance there is?
%     I think that would be useful: to tell people that approximately
%     equal parts of the error bars come from instrument noise and from
%     sample variance.
% LP: I think the note is useful - tell me what you think.

{\footnotesize
\bigskip
Table~1. --- Flight~2 Angular Power Spectrum.
The band powers $\delta T_\l = [\l(\l+1)C_\l/2\pi]^{1/2}$ 
have window functions whose mean and 
rms width are given by $\l_{eff}$ and $\Delta\l$.
The errors $\delta T$ include both noise and sample variance, 
and are uncorrelated for the two Ka-points.
%{\bf Comment on fraction that's sample variance.}
\smallskip
%  \tabcaption{The Angular Power Spectrum}
\begin{center}
\begin{tabular}{crcl}
\hline
\hline
\multicolumn{1}{l}{Band}             	  &
\multicolumn{1}{c}{$\l_{eff}$}          & 
\multicolumn{1}{c}{$\Delta \l $}  & 
\multicolumn{1}{c}{$\delta T$ [$\mu$K]}    \\
\hline
Ka	&91	&47  &$46^{+10}_{-12}$\\
Ka	&145	&64  &$63^{+10}_{-12}$\\
Q	&125	&67  	&$56^{+5}_{-6}$\\
\hline
\end{tabular}
\end{center}
\smallskip
}
\bigskip

% sky coverage is 79.57 square degrees using 0.1x0.1 degree pixels

\medskip
\vbox{
% \vskip-4.6cm
\centerline{\epsfxsize=9.0cm\epsfbox{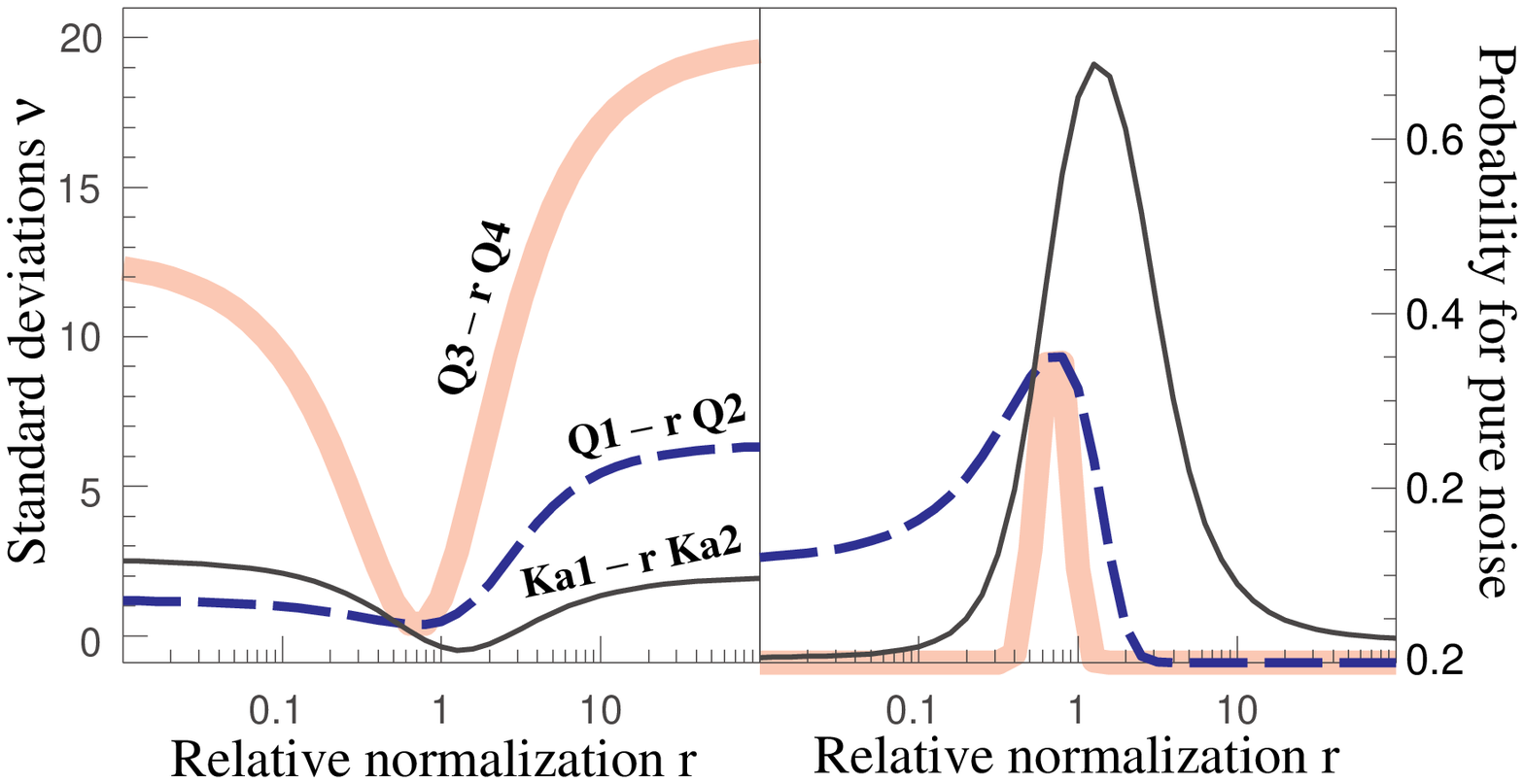}}
% \vskip-1.0cm
\figcaption[null.ps]{Evidence for real sky signal in the maps.
The left panel shows the signal-to-noise ratio
$\nu$ in difference maps weighted by $r$ (see text) for
Ka1 vs.~Ka2 (solid black), Q1 vs.~Q2 (dashed) and Q3 vs.~Q4 (grey).
The right panel shows the corresponding probability that the 
difference map contains noise only.
This figure shows that all maps except Q1 contain significant
detection of anisotropy and that the signal is common to both maps of each
polarization pair.
% For each pair of maps, the left panel shows the signal-to-noise ratio
% $\nu$ in a difference map weighted by $r$ (see text), while the right
% panel shows the probability that the difference map contains noise only.
% This figure demonstrates that each individual map contains a significant
% detection of anisotropy and that the signal is common to both maps of a
% polarization pair.
\label{fig:null}}}
\smallskip

% get exact sky coverage

\acknowledgments

We are grateful for the invaluable support of the NSBF and we thank Eric
Torbet for his contributions. 
This work was supported by
a David \& Lucile Packard Foundation Fellowship (to LP),
a Cottrell Award from Research Corporation,
an NSF NYI award, 
NSF grants PHY-9222952 and PHY-9600015, 
NASA grant NAG5-6034 and
Hubble Fellowships 
HF-01044.01$-$93A (to TH) and
HF-01084.01$-$96A (to MT)
from by STScI, operated by AURA, 
Inc. under NASA contract NAS5-26555.

\goodbreak

%{\bf LEFT TO DO 
%1) Check that statements on errors and significance are correct
%2) Put in correct astro-ph numbers to D98 and dO98 on submission.}
%\vfill\eject

\end{document}